\documentclass[11pt]{iopart}

\usepackage{iopams}
\usepackage{epsfig}
\def\dd{\displaystyle}

\begin{document}

\title{Dirichlet Casimir Energy for a Scalar Field in a Sphere: An Alternative Method}

\author{M.A. Valuyan, and S.S. Gousheh}
\address{ Department of Physics, Shahid Beheshti University, Evin,
Tehran 19839, Iran }

\ead{m-valuyan@sbu.ac.ir}

\begin{abstract}
In this paper we compute the leading order of the Casimir energy for
a free massless scalar field confined in a sphere in three spatial
dimensions, with the Dirichlet boundary condition. When one
tabulates all of the reported values of the Casimir energies for two
closed geometries, cubical and spherical, in different space-time
dimensions and with different boundary conditions, one observes a
complicated pattern of signs. This pattern shows that the Casimir
energy depends crucially on the details of the geometry, the number
of the spatial dimensions, and the boundary conditions. The
dependence of the \emph{sign} of the Casimir energy on the details
of the geometry, for a fixed spatial dimensions and boundary
conditions has been a surprise to us and this is our main motivation
for doing the calculations presented in this paper. Moreover, all of
the calculations for spherical geometries include the use of
numerical methods combined with intricate analytic continuations to
handle many different sorts of divergences which naturally appear in
this category of problems. The presence of divergences is always a
source of concern about the accuracy of the numerical results. Our
approach also includes numerical methods, and is based on Boyer's
method for calculating the electromagnetic Casimir energy in a
perfectly conducting sphere. This method, however, requires the
least amount of analytic continuations. The value that we obtain
confirms the previously established result.
\end{abstract}

\section{Introduction}
The Casimir effect is one of the most interesting manifestations of
the nontrivial structure of the vacuum state in quantum field
theory. This effect appears when nontrivial boundary conditions or
background fields are present (\emph{e.g.} solitons). In 1948
Casimir predicted the existence of this effect as an attractive
force between two infinite parallel uncharged perfectly conducting
plates in vacuum \cite{h.b.g.1,h.b.g.2} (for a general review on the
Casimir effect, see
Refs.\,\cite{mostepanenko.,milton.,greiner.,milton.book.,lecture.note.elizalde}).
This effect was subsequently observed experimentally by Sparnaay in
1958 \cite{sparnaay}. Up to now, most experiments on the measurement
of the Casimir forces have been performed with parallel
plates\,\cite{Bressi.} or with a sphere in front of a
plane\,\cite{Lamoreaux.,decca1.,decca2.,geyer.}. Furthermore, the
configuration of two eccentric cylinders have both experimental and
theoretical interest\,\cite{dalvit.Europhys.}. The majority of the
investigations related to the Casimir effect concern the calculation
of this energy or the resulting forces for different fields in
different geometries with different boundary conditions
imposed\,\cite{Kirsten.book.2}. These cases include parallel
plates\,\cite{h.b.g.1,h.b.g.2},
cubes\,\cite{cube.article.1,cube.article.2,cube.article.3,cube.article.4,cube.article.5,Lukosz.},
cylinders\,\cite{Neto.,Mazzitelli.,dalvit1.,VVNesterenko1.,VVNesterenko2.,Francisco.,Gosdzinsky.,Rodrigues.},
and spherical
geometries\,\cite{Kirsten.book.2,prd.56.4896.,Elizalde.JPAmg.,Bender.}.
This effect has even been investigated in connection with the
properties of the space-time with extra
dimensions\,\cite{lecture.note.elizalde,Poppenhaeger.,Carlos.}, and
nowadays it is known that the Casimir energy depends strongly on the
geometry of the space-time and on the boundary conditions
imposed\,\cite{saharian1.,saharian2.}. In fact, an interesting
question is the determination of the conditions under which the
forces acting on the boundaries for closed geometries are attractive
or repulsive in arbitrary spatial
dimensions\,\cite{caruso...,RMCavalcanti.,AEdery.,HAlnes.,MPHertzberg.,XLi.}.
\par
The Casimir effect has many applications in different branches of
physics. Perhaps the first extensive study of the Casimir effect was
in Particle Physics and in connection with the development of the
bag model of
hadrons\,\cite{milton.book.,prd.56.4896.,chodos.,RKBhaduri.,sphere.article1.,sphere.article2.,sphere.article3.,sphere.article4.,sphere.article5.}.
One of the most fascinating and open problems of theoretical physics
is the cosmological
constant\,\cite{grattini.,Elizalde.PLB.,Elizalde.JPhysA.,FBauer.,cheng.hong.bo.}.
This constant is also a candidate for the dark
energy\,\cite{Elizalde.PLB.,FBauer.,plb.1}, and the Casimir energy
has been studied in this
connection\,\cite{Elizalde.PLB.,FBauer.,j.math.elizalde.}. Moreover,
the presence of the Casimir forces in many different phenomena of
condensed matter and laser physics have been established both
theoretically and
experimentally\,\cite{many1.,many2.,many3.,many4.}. In particular,
the study of the Casimir effect for massless scalar fields is not
only of theoretical interest but also has direct relevance to
physical systems such as Bose-Einstein
condensates\,\cite{bose.cite1.,bose.cite2.,bose.cite3.}.
\par
Calculations of the Casimir energy in spherically symmetric
configurations have attracted the interest of physicists for many
years\,\cite{prd.56.4896.,Elizalde.JPAmg.,Bender.,sphere.article1.,sphere.article2.,sphere.article3.,sphere.article4.,sphere.article5.}.
As Boyer\,\cite{boyer.} first showed, the Casimir pressure exerted
by the electromagnetic\,(EM) field on the walls of a perfectly
conducting spherical vessel is repulsive (see
also\,\cite{milton.raad.}). The reported Casimir energy and pressure
of a scalar field confined in a spherical vessel with Dirichlet
boundary condition is also reported to be
positive\,\cite{positive1.,positive2.,positive3.,cognola.,nestrenko.,Bowers.}.
Comparison between the reported values of the Casimir energy for
closed geometries show that these values depend crucially on three
major factors: first the details of the geometries (e.g. spheres or
cubes), second the number of space dimensions, and third the
boundary conditions. To be very concrete we have collected all of
the reported results for various cases and display them in
Table\,(\ref{tab.compare.}). We should mention that all of the
results obtained for cubical geometry are exact, while the ones for
spherical geometries are obtained approximately by various numerical
methods.
\begin{table}[th]
 \hspace{2cm}\begin{tabular}{||c|c|c|c|c||}
  \hline
     \small{Dimension} & Field      &   \small{B.C.s}&    \small{Cube}          & \small{Sphere}    \\\hline\hline
  \small{2}& Scalar  & \small{$\mathcal{D}$}&\small{+0.041}      & \small{+0.000672-0.003906/s}                            \\
  \small{2} & Scalar  & \small{$\mathcal{N}$}&\small{-0.22}      & \small{-0.183123-0.019531/s}                               \\
  \small{2} & EM & \small{Conductor}&\small{-0.22}      & \small{-0.183123-0.019531/s}                          \\\hline
  \small{3} & Scalar  & \small{$\mathcal{D}$}&\small{-0.016}  & \small{+0.002819}                    \\
  \small{3} & Scalar  & \small{$\mathcal{N}$}&\small{-0.29}  & \small{-0.223458}                  \\
  \small{3} & EM      &\small{Conductor}&\small{+0.092}&\small{+0.046200}     \\\hline
  \small{4} & Scalar   & \small{$\mathcal{D}$}&\small{+0.0061}  & \small{-0.000655+0.000267/s}                    \\
  \small{4} & Scalar  & \small{$\mathcal{N}$}&\small{-0.33}  & \small{-0.260872-0.044716/s}                    \\
  \small{4} & EM  & \small{Conductor}&\small{-0.044}  & \small{-0.197834-0.033768/s}                  \\\hline
  \small{5} & Scalar  & \small{$\mathcal{D}$}&\small{-0.0025}  & \small{-0.000288}                  \\
   \small{5} & Scalar  & \small{$\mathcal{N}$}&\small{-0.37}  & \small{-0.270281}                  \\
    \small{5} & EM  & \small{Conductor}&\small{+0.021}  & \small{-0.006362}                  \\\hline
 \end{tabular}\caption{\label{tab.compare.} \small
The Casimir energies for the EM field (with boundary condition
consistent with a perfectly conducting cavity denoted by
``Conductor") and massless scalar fields (with Dirichlet and Neumann
boundary conditions denoted by ``$\mathcal{D}$" and
``$\mathcal{N}$", respectively). The values are displayed for the
space dimensions $D=\{2,3,4,5\}$, for cubical and spherical
geometries. The quantity $s$ is a regularization factor which should
be taken to zero. The values are displayed for a cube with unit
volume and a sphere with radius $a=1$, all in units $\hbar c=1$.}
  \label{geometry}
\end{table}
As one can see from the Table\,(\ref{tab.compare.}), for the
spherical cases in even spatial dimensions, there always remains an
unresolved divergent factor. For the case of massless scalar fields
with the Neumann boundary condition, the Casimir energy is always
negative regardless of the details of the geometry and the number of
space dimensions. Therefore, one might conclude that there are no
surprises there. However, for the case of massless scalar fields
with Dirichlet boundary conditions there is a sign factor $(-1)^D$
for the case of cubes\,\cite{x.li.}, and
$(-1)^{\mbox{\footnotesize{Floor}}(D/2-1)}$ for the case of
spheres\,\cite{cognola.}, where $D$ denotes the number of space
dimensions. For the case of EM field inside a perfectly conductor
there is a sign factor $(-1)^{D+1}$ for the case of cube, and no
obvious sign factor for the case of sphere.
\par
Now we concentrate on the case of three spatial dimensions. At a
first glance, the reported results for the case of a massless scalar
field confined in a spherical geometry with Dirichlet boundary
condition seem anomalous because it is the only case for which the
two geometries do not have the same sign of the Casimir energy, and
this has caused a controversy in the literature. As far as this
controversy is concerned, L.A. Manzoni and W.F.
Wreszinski\,\cite{manzoni.}, try to justify these results by showing
that the Casimir forces in both cases are repulsive. It has been
claimed that for the EM case, the deformation of a spherical shell
of radius $a$ into a cubical shell of length $L$ with $L\approx 2a$,
should not change the sign and approximately the magnitude of the
Casimir energy, when the boundary conditions are
unchanged\,\cite{Lukosz.,wolf.,peterson.hansson.}. We believe that
their claim is reasonable. However, it is inconsistent with the
results for the EM field in five spatial dimensions and the massless
scalar field with Dirichlet boundary conditions in three and four
spatial dimensions. Therefore, it is worth studying this problem in
more details, and here we concentrate on the three dimensional case.

\par Although, the Casimir energy for the case of the cube is exactly
solvable\,\cite{wolf.}, the presence of divergences inherent in
these sorts of problems require regularization and analytic
continuation methods\,\cite{Reuter.}, and this has been a source of
criticism for this calculation\,\cite{cavalcanti2.,hertzberg.}.
Since the Casimir energy for the case of the sphere is not exactly
solvable, it requires the utilization of numerical methods. The use
of numerical methods for problems which include divergences could be
a source of even greater concerns about the accuracy of the final
results. When the problem is plagued with multitude of divergences,
the numerical methods should include delicate and carefully planed
regularization and possibly analytical continuation techniques. The
problem addressed in this paper is in this category, and we
calculate this case using an alternative method which requires the
least amount of regularization and analytic continuation schemes.
%
%
\par There are three general methods for calculating the Casimir
energy for the spherical case, which as mentioned earlier all
include numerical parts. The first method is the Green function
formalism\,\cite{Bender.}. In this method one encounters an infinite
sum of integrals over the modified Bessel functions. It can be shown
that each of the integrals converges and the sum can be calculated
numerically by using the asymptotic behavior of the modified Bessel
functions for all real D. The plot of the Casimir pressure as a
function of continuously variable D show that the result is infinite
for even dimensions\,\cite{Bender.}. The second method is the zeta
function technique. For example, G. Cognola \textit{et
al}\,\cite{cognola.} use the zeta function technique to obtain the
Casimir energies for spherical symmetric cavities, for different
boundary conditions, and fields in D space dimensions. The third
method is the direct mode summation using contour
integration\,\cite{nestrenko.}. This method is based on the direct
summation of the frequencies and the main tool employed is the
Cauchy theorem. This employment lets the authors sum the zeros of
the Bessel functions more easily.

The method we use in this paper is based on the Boyer's method which
was used for the EM field in a sphere. An important part of the
Boyer's method is to confine the system inside a similarly shaped
but larger shell, and then to subtract the energies of two
configurations with the same size outer shells and different size
inner shells. In this subtraction scheme, most of the infinities
automatically cancel without any need to use analytic continuations.
As a matter of fact in this method, the Casimir energy is simply the
work done on the system in deforming the initial configuration to
the one under consideration. Therefore, the quantity just defined
has an obvious physical interpretation. One can then let the radii
of the outer shells and the second inner shell go to infinity. We
shall henceforth refer to this subtraction scheme as the Boyer
Subtraction Scheme (BSS). We recently used this method to directly
compute the lowest order Casimir energy for the EM field inside a
rectangular waveguide Ref.\,\cite{madad.}, and the lowest order
radiative corrections for the Casimir energy for a scalar field for
the parallel plate problem in various space-time dimensions
\,\cite{reza1.,reza2.}.
\par
In this paper we use BSS for a massless scalar field confined in a
sphere with the Dirichlet boundary condition. The analytic part of
our calculation is analogous to the one used by Boyer for the TE
modes, and is done in section 2. In section 3, we present the final
part of our calculations which, similar to the Boyer's paper, is
numerical.  However, our numerical calculations differ from that
paper. We introduce a method to extrapolate the sum of all the zero
point energies for a given value of the angular momentum $\ell$. In
order to proceed with the summation over $\ell$, we first introduce
a second extrapolation method to obtain the optimal form of the
summand. We then sum over all $\ell$ by using the zeta function
analytic continuation technique to remove the infinities and extract
the finite part. Our final result agrees with the established value
obtained
in\,\cite{positive1.,positive2.,positive3.,nestrenko.,Bowers.}. In
section 4, we summarize and discuss our results.

\section{Analytic Part of the Calculation of Casimir Energy in a Sphere}
In this section we setup all the physical and mathematical machinery
for the calculation of the Casimir energy for a real massless scalar
field with Dirichlet boundary condition for a spherical geometry.
The mathematical part introduced in this section parallels closely
the one due to Boyer\,\cite{boyer.} for the computation of the
Casimir energy for the EM field inside a perfectly conducting
sphere. The similarity between our calculation and that of Boyer
stems from the fact that the expressions for the Casimir energy for
the TE mode of the EM field is similar to the massless scalar field
with Dirichlet boundary condition\,\cite{milton.book.}. The
important difference is that the term with $\ell =0$ is allowed in
the latter case. Moreover, there is no analogue of the TM mode in
our problem due to choice of boundary condition. As we shall see,
this forces us to encounter some new divergences. A close
examination of Boyer's work show that when both modes are present
these divergences cancel. The final expression obtained in this
section cannot be solved analytically, and we solve it numerically
in the next section.

\begin{figure}[th] \hspace{4cm}\includegraphics[width=7.3cm]{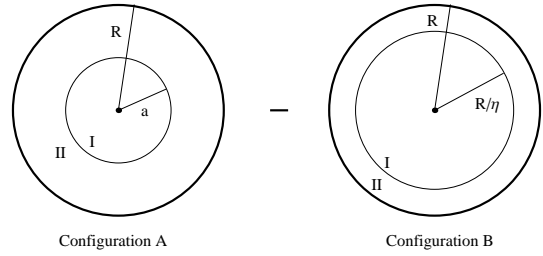}\caption{\label{two.spheres.fig.} \small
  Left figure shows two concentric spheres with radii $a$ and $R$ (configuration `A').
  The right figure shows a similar configuration with different radius for the inner sphere
  (configuration `B'). To calculate the Casimir energy, the zero point energies of these two
  spherical configurations are to be subtracted according to the Eq.\,(\ref{ECas.Def.}).
  The label `I' refers to the inner sphere and `II' the outer annular region. In the final
  step, $R\to\infty$, while $\eta$ is kept fixed.}
  \label{geometry}
\end{figure}

The Casimir energy is defined as the difference between the sum of
the zero point energies of all the modes in the bounded region
considered, and the free case. In order to find all of the modes
inside a sphere we have to solve the Klein-Gordon equation for a
massless scalar field, with the appropriate boundary condition. Due
to spherical symmetry, each energy level $E_{\ell}$ will be
$(2\ell+1)$\,fold degenerate. Therefore we have the following
expression for the total zero point energy,
\begin{eqnarray}\label{zero.point.}
  E=\sum\limits_{\ell = 0}^{\infty}\left(2\ell+1\right)
  \sum\limits_{s = 1}^{\infty}\frac{\hbar\omega_{\ell,s}}{2}=\frac{\hbar c}{2}
  \sum\limits_{\ell = 0}^{\infty}\left(2\ell+1\right)
  \sum\limits_{s = 1}^{\infty}k_{\ell,s},
\end{eqnarray}
where $\omega_{\ell,s}$ and $k_{\ell,s}$ are the normal modes
frequencies and wave-vectors, respectively, and the index $s$ refers
to the root number for a given value of $\ell$. Since there are an
infinite number of normal modes of increasingly high frequency, this
energy is infinite, as usual. However, as mentioned before
$E_{\mbox{\footnotesize Cas.}}$ is obtained by the subtraction of
this zero-point energy from the analogous one for the free case. An
essential part of the Boyer's method is to enclose this sphere
inside a larger one. Then a similar configuration is considered but
with a different radius for the inner sphere (see
Fig.\,(\ref{two.spheres.fig.})). Then the zero point energies of
these two configurations are subtracted from each other, and many of
the infinities cancel each other out. Then the radii of all the
spheres except the original inner sphere (the one with radius `$a$')
is taken to infinity.
Therefore our expression for the Casimir energy is given in Eq.\,(\ref{ECas.Def.}),
\begin{eqnarray}\label{ECas.Def.}
  E_{\mbox{\footnotesize Cas.}}= \lim_{R\to \infty}\left[(E^{A}_{I}+E^{A}_{II})-(E^{B}_{I}+E^{B}_{II})\right].
\end{eqnarray}
Using Eqs.\,(\ref{zero.point.},\ref{ECas.Def.}) we have,
\begin{eqnarray}\label{ECas.replace.}\hspace{-2.5cm}
  \small{E_{\mbox{\footnotesize Cas.}}}=\small{\lim\limits_{R\to \infty}\lim\limits_{\lambda\to 0}\frac{\hbar c}{2}
  \sum\limits_{\ell=0}^{\infty}\left(2\ell+1\right)}\\ \hspace{-0.5cm}
 \times\Bigg[\sum\limits_{s=1}^{\infty}k_{\ell,s}\left(a\right)\,g\left(\lambda
  k_{\ell,s}\left(a\right)\right)
  +\sum\limits_{S=1}^{\infty}K_{\ell,S}\left(a,R\right)\,g\left(\lambda K_{\ell,S}\left(a,R\right)\right) -\{a\to R/\eta\}\Bigg],\nonumber
\end{eqnarray}
where $k_{\ell,s}(a)$ and $K_{\ell,S}\left(a,R\right)$ are the
wave-vectors for the inner sphere and annular region, respectively,
and $g(\lambda k_{\ell,s}(a))=e^{-\lambda k_{\ell,s}\left(a\right)}$
is a convergence factor which eventually goes to one as $\lambda\to
0$. Moreover the index $S$ is defined analogously to $s$, but for
the annular region.
\par
The imposition of the boundary condition is explained in
\ref{Calculation.1}, where the relationships between the root
indices, $s$ and $S$, and the wave-vectors
$k_{\ell,s}\left(a\right)$ and $K_{\ell,S}\left(a,R\right)$, are
obtained. Using the results obtained in \ref{Calculation.1} and the
Euler-Maclaurin Summation Formula\,(EMSF), we calculate the limit $R
\to \infty$ of terms in Eq.\,(\ref{ECas.replace.}) coming from the
annular regions in \ref{Calculation}. There we show how in this
limit, most of the infinities coming from these regions cancel due
to BSS, and all of the remaining finite terms approach zero. The
final result is that the two indicated sums turn into integrals with
no remainder. As explained below in more details, the two integrals
eventually cancel the infinities coming from their respective inner
spheres, and their residual infinities cancel each other. We finally
obtain,
\begin{eqnarray}\label{ecas.1}\hspace{-2.5cm}\vspace{-0.4cm}
  E_{\mbox{\footnotesize Cas.}}=\mathop{\lim}\limits_{R\to
  \infty}\mathop{\lim}\limits_{\lambda\to 0}\frac{\hbar
  c}{2}\sum\limits_{\ell=0}^{\infty}(2\ell+1)\Bigg[\sum\limits_{s=1}^{\infty}k_{\ell,s}(a)g(\lambda
  k_{\ell,s}(a))\nonumber\\ \hspace{3.3cm} +\int_{S=1}^{\infty}dS\, K_{\ell,S}(a,R)g(\lambda
  K_{\ell,S}(a,R)) -\{a\rightarrow R/\eta\}\Bigg],
\end{eqnarray}
where the continuous variable $S$ is associated with the discrete
root number $S_{\ell}(a,K)$ and appears when we use EMSF. The above
equation is similar to the expression obtained by Boyer for the TE
mode, with the only difference being the inclusion of $\ell =0$ in
this case. It is obvious from their definitions that $K_{\ell,S}(a)$
and $S_{\ell}(a,K)$ are in one to one correspondence. Therefore we
can change the variable of integration from $S$ to the continuous
version of the wave-vectors, which we shall denote by $K$ for
brevity of notation, and using Eq.\,(\ref{s and S eqs.}) we have,
\begin{eqnarray}\label{ecas.2}\hspace{-2.5cm}\vspace{-0.3cm}
  E_{\mbox{\footnotesize Cas.}}=\small{\mathop{\lim}\limits_{R\to
  \infty}\mathop{\lim}\limits_{\lambda\to 0}\frac{\hbar
  c}{2}\sum\limits_{\ell=0}^{\infty}(2\ell+1)\Bigg[\sum\limits_{s=1}^{\infty}k_{\ell,s}(a)g(\lambda
  k_{\ell,s}(a))-\dd\int_{K=K_{\ell,1}(a,R)}^{\infty}\hspace{-0.5cm}dK\,
  \frac{ds_{\ell}(a,K)}{dK}K\,g(\lambda K)} \nonumber\\ \hspace{3.2cm}
  +\int_{K=K_{\ell,1}(a,R)}^{\infty}dK\,
  \frac{ds_{\ell}(R,K)}{dK}K\,g(\lambda K)-\{a\rightarrow
  R/\eta\}\Bigg].\hspace{0.4cm}
\end{eqnarray}
Rewriting Eq.\,(\ref{ecas.2}) and rearranging terms we have,
\begin{eqnarray}\label{ecas.3}\hspace{-2.5cm}
  E_{\mbox{\footnotesize Cas.}}=\mathop{\lim}\limits_{R\to
  \infty}\mathop{\lim}\limits_{\lambda\to 0}\frac{\hbar
  c}{2}\sum\limits_{\ell=0}^{\infty}(2\ell+1)\Bigg\{\Bigg[\sum\limits_{s=1}^{\infty}k_{\ell,s}(a)g(\lambda
  k_{\ell,s}(a))-\int_{K=K_{\ell,1}(a,R)}^{\infty}\hspace{-0.5cm}dK\,
  \frac{ds_{\ell}(a,K)}{dK}K\,g(\lambda K)
  \nonumber\\ \hspace{1.1cm}
  \hspace{2cm}-\{a\rightarrow R/\eta\}\Bigg]-\int_{K=K_{\ell,1}(R/\eta,R)}^{K=K_{\ell,1}(a,R)}dK\frac{ds_{\ell}(R,K)}{dK}K\,g(\lambda
  K)\Bigg\}.
\end{eqnarray}
As shown in \ref{Calculation},
Eqs.\,(\ref{K(a)-K(R/eta.1},\ref{K(a)-K(R/eta.2}), in the limit
$R\to \infty$, the wave-vectors $K_{\ell,1}(a,R)$ and
$K_{\ell,1}(R/\eta,R)$ not only decrease exponentially in $\ell$,
but also go to zero as $1/R$. So, the difference between the upper
and lower limits of the last integral approaches zero and since the
integrand is not singular, there is no contribution coming from this
term to $E_{\mbox{\footnotesize Cas.}}$. Also, the lower limits of
the remaining integrals can be extended to $\small{K=0}$.

We can use the simplest form of the EMSF to make the first term in
Eq.\,(\ref{ecas.3}) more amenable to computation,
\begin{eqnarray}\label{euler.mac.sum.f1}
  &&\hspace{-2.5cm}\sum\limits_{s=1}^{\infty}k_{\ell,s}(a)g(\lambda
  k_{\ell,s}(a))=\int_{s=1}^{\infty}ds\, k_{\ell,s}(a)\,g(\lambda
  k_{\ell,s}(a))+\frac{1}{2}[k_{\ell,1}(a)g(\lambda k_{\ell,1}(a))]
  \nonumber\\ &&\hspace{1.5cm}+\int_{s=1}^{\infty}ds \Big(s-[s]-\frac{1}{2}\Big)\frac{d}{ds}[k_{\ell,s}(a)\,g(\lambda
  k_{\ell,s}(a))],
\end{eqnarray}
where $[s_{\ell}(a,k)]$ denotes the Floor Function. The one to one
correspondence of $s_{\ell}(a,k)$ and $k_{\ell}(a,s)$ is  analogous
to the previous case which was for the annular regions. Therefore,
we can again change the variable of the integration from $s$ to $k$,
and by adding and subtracting appropriate terms we can extend all of
the lower limits of the integrals to zero. We obtain:
\begin{eqnarray}\label{changing.var.}\hspace{-2cm}
\sum\limits_{s=1}^{\infty}k_{\ell,s}(a)g(\lambda
  k_{\ell,s}(a))=\int_{k=0}^{\infty}dk\frac{ds_{\ell}(a,k)}{dk}\, k\,g(\lambda
  k)-\int_{k=0}^{k=k_{\ell,1}(a)}dk\frac{ds_{\ell}(a,k)}{dk}\, k\,g(\lambda k)
  \nonumber\\ \hspace{+1.8cm}+\frac{1}{2}\big[k_{\ell,1}(a)g(\lambda k_{\ell,1}(a))\big]+\int_{k=0}^{\infty}dk
  \Big(s-[s]-\frac{1}{2}\Big)\frac{d}{dk}[k\,g(\lambda k)]
  \nonumber\\ \hspace{+1.8cm}-\int_{k=0}^{k=k_{\ell,1}(a)}dk  \Big(s-[s]-\frac{1}{2}\Big)\frac{d}{dk}[k\,g(\lambda k)].
\end{eqnarray}
The last term of Eq.\,(\ref{changing.var.}) can be simplified by
noting that the Floor Function $[s_{\ell}(a,k)]=0$ in the indicated
domain, and integration by parts finally yields,
\begin{eqnarray}\label{3rd.term.}\hspace{-2.5cm}
  \int_{k=0}^{k=k_{\ell,1}(a)}dk
  \Big(s_{\ell}(a,k)-[s_{\ell}(a,k)]-\frac{1}{2}\Big)\frac{d}{dk}[k\,g(\lambda
  k)]=\int_{k=0}^{k=k_{\ell,1}(a)}dk\,\Big(s_{\ell}(a,k)-\frac{1}{2}\Big)\frac{d}{dk}[k\,g(\lambda
  k)]\nonumber\\ \hspace{3.5cm}
  =\frac{1}{2}k_{\ell,1}(a)g(\lambda
  k_{\ell,1}(a))-\int_{k=0}^{k=k_{\ell,1}(a)}dk\,
  \frac{s_{\ell}(a,k)}{dk}k\,g(\lambda k).
\end{eqnarray}
Finally Eq.\,(\ref{changing.var.}) can be simplified to,
\begin{eqnarray}\label{euler.mac.sum.f2}\begin{array}{ll}\hspace{-2.8cm}
  \sum\limits_{s=1}^{\infty}k_{\ell,s}(a)g(\lambda k_{\ell,s}(a))=\hspace{-0.2cm}
  \dd\int_{k=0}^{\infty}\hspace{-0.2cm}dk\,\Bigg[
  \dd\frac{ds_{\ell}(a,k)}{dk}k\,g(\lambda k)+
  \Big(s_{\ell}(a,k)-[s_{\ell}(a,k)]-\frac{1}{2}\Big)\frac{d}{dk}[k\,g(\lambda
  k)]\Bigg].\hspace{0.5cm}
\end{array}\end{eqnarray}
Upon substituting the expression displayed in
Eq.\,(\ref{euler.mac.sum.f2}) into Eq.\,(\ref{ecas.3}), the
divergent terms cancel each other (the first term of
Eq.\,(\ref{euler.mac.sum.f2}) and the second term of
Eq.\,(\ref{ecas.3})). Analogous cancelation occurs for the other
regions. All of the remaining terms are finite. Moreover, we can
make the integrals dimensionless by making appropriate changes of
variables, such as $x=ak$. We finally obtain
\begin{eqnarray}\label{final. cas.}\hspace{-2.5cm}
  E_{\mbox{\footnotesize Cas.}}=\lim\limits_{R\to \infty}\lim\limits_{\lambda\to
  0}\frac{\hbar
  c}{2}\Bigg[\sum\limits_{\ell=0}^{\infty}(2\ell+1)\Bigg(\frac{1}{a}\int_{x=0}^{\infty}dx
  \Big(s_{\ell}(x)-[s_{\ell}(x)]-\small{\frac{1}{2}}\Big)\frac{d}{dx}[x\,g((\lambda/a)x)]\Bigg)\nonumber\\ \hspace{1.5cm}-\{a\rightarrow
  R/\eta\}\Bigg].
\end{eqnarray}
This expression is analogous to the one for the Casimir energy for
the TE mode obtained by Boyer for the same geometry, except that
$\ell=0$ is included in our expression. This integral cannot be done
analytically and we compute it numerically in the next section.

\section{Numerical Evaluation of the Casimir Energy}
\begin{figure}
     \hspace{2cm}\includegraphics[width=6cm]{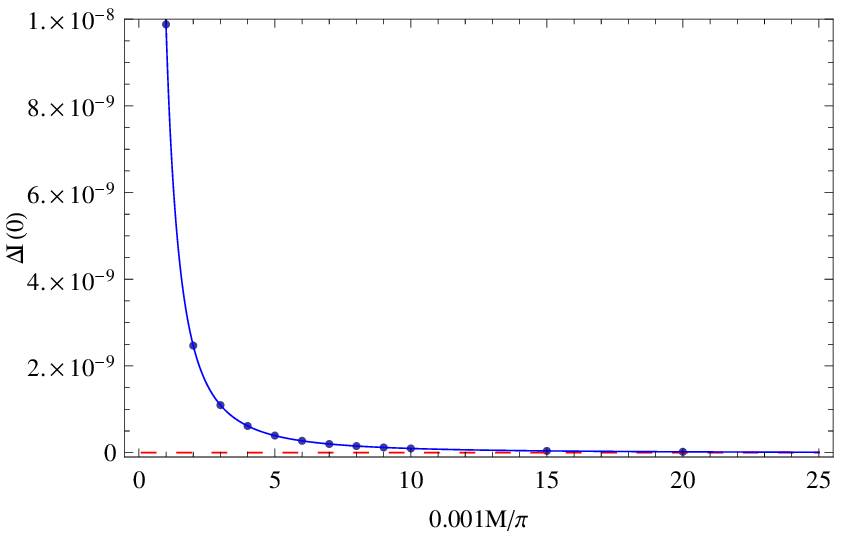}    \includegraphics[width=6cm]{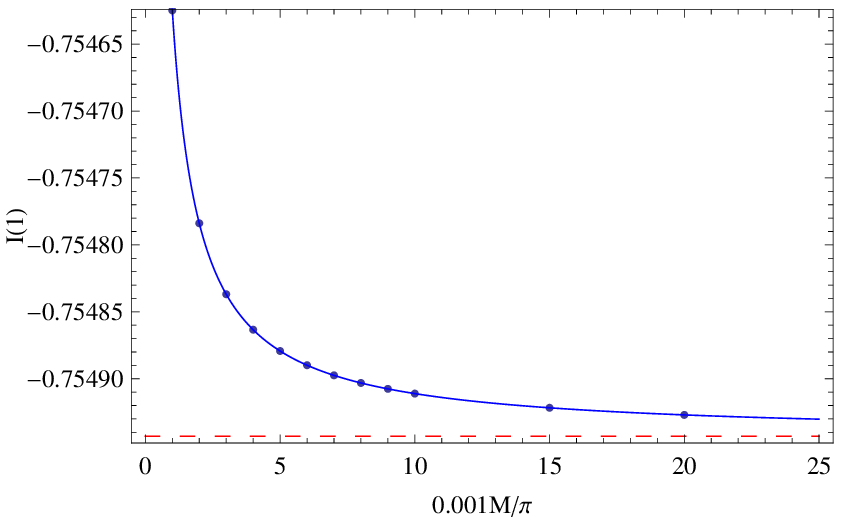}
\caption{\label{figs.assym.M.}   In the left figure, we plot the
difference between our numerical results for $\ell=0$ (the data
points and their best fit shown using solid line), and the
theoretical value $E_{\mbox{\footnotesize
Cas.}}=\frac{-\pi}{12}$\,(in units $\frac{\hbar c}{2a}=1$) as a
function of the cutoff $M$, as explained in the text. Note that, our
numerical values asymptotically approach the theoretical value. In
the right part of the figure we show our numerical results for the
case $\ell=1$ (the data points and their best fit shown using solid
line). For both figures the asymptotic values are obtained by
fitting a sum of polynomials\,($r^{n}$) with $n=\{-5,...,0\}$ and
are shown using dashed lines.  }
\end{figure}

Analytic calculation of Eq.\,(\ref{final. cas.}) seems to be
impossible. Therefore, we resort to a numerical method. We compute
the integrals in  Eq.\,(\ref{final. cas.}) for each value of $\ell$
separately, and then sum the results as indicated in that equation.
However, the numerical method used by Boyer cannot be employed,
since  in our problem we do not have the luxury of cancelation of
infinities between the TE and TM modes and the constancy of their
sum. As is apparent from Eq.\,(\ref{final. cas.}), the integrand for
each $\ell$ has an infinite number of discontinuities due to the
presence of the Floor Function in that expression. After determining
the precise position of the jumps, the integrations are done
separately for all parts and then all of the results are summed. The
integration is over the continuous version of the wave number, which
extends to infinity. In order to accomplish this numerically we
compute the integral up to a cutoff $M$. For fixed $\ell$, the
quantity $M$ can equivalently be thought of as a cutoff over the
root numbers $s$. Then we compute this integral for a series of
values of $M$. The results of the numerical integration, indicated
by $I(\ell,M)$, are plotted as a function of $M$. Then by fitting a
polynomial function to this plot, the asymptotic value can be easily
obtained, and gives us the infinite $M$ limit.  On the other hand we
have to take the limit $\lambda\to 0$ as indicated in
Eq.\,(\ref{final. cas.}). An attempt to optimize the accuracy of our
results as a function of the relationship between $M$ and $\lambda$
has revealed that the best choice is $\lambda=1/M$. In
Fig.\,(\ref{figs.assym.M.}) we display the computational technique
just described for the cases $\ell=\{0,1\}$.
The case $\ell=0$ also serves a secondary purpose, which is a
crucial part for the rest of our numerical analysis. For this case
the frequencies are proportional to integer multiples of $\pi$ and
the original expression for the Casimir energy,
Eq.\,(\ref{zero.point.}), can be computed directly as follows:
\begin{eqnarray}\label{exact.l=0.}\hspace{-1.2cm}
E= \frac{\hbar}{2}\sum\limits_{s = 1}^{\infty}\omega_{0,s}=\frac{\hbar c}{2}\sum\limits_{s = 1}^{\infty}k_{0,s}=\frac{\hbar
  c}{2}\sum\limits_{s=1}^{\infty}\frac{s\pi}{a}=\frac{\hbar c \pi}{2a}\zeta(-1)\rightarrow \small{\frac{\hbar c
  \pi}{2a}\Big(\frac{-1}{12}\Big)}.
\end{eqnarray}
Note that in the last step indicated by arrow we have used the
analytic continuation of the zeta
function\,\cite{lecture.note.elizalde,Elizalde.book.1}. In the left
part of Fig.\,(\ref{figs.assym.M.}) we have plotted the difference
between our numerical results for various values of the cutoff $M$,
for the case $\ell=0$, and the theoretical value
$\frac{-\pi}{12}$\,(in units $\frac{\hbar c}{2a}=1$). As it is
apparent from this figure, the asymptotic values of our numerical
results exactly matches the theoretical one. This clearly shows the
consistency between our numerical methods and the theoretical
results obtained by analytical continuation of the zeta function. In
the right part of Fig.\,(\ref{figs.assym.M.}), we plot the actual
numerical values of the Casimir energy for $\ell=1$, since
theoretical values do not exist in this case. In
Table\,(\ref{tab.3}) all the results obtained for various values of
$\ell$, up to $\ell=21$, are shown for various values of the cutoff.
Note that the values indicated in the last column are the asymptotic
values which correspond to the cases in which the cutoffs have been
sent to infinity.
\begin{table}[th]
 \hspace{0.5cm}\begin{tabular}{||c|c|c|c|c|c|c||}
  \hline
  $\ell$&  \small{$M=1000\pi$ }& \small{$M=5000\pi$}&  \small{$M=10000\pi$} &  \small{$M=15000\pi$} &  \small{$M=20000\pi$} &   \small{$M\to \infty$} \\\hline\hline \vspace{-0.15cm}
  \small{0}    &  \small{-0.26179}  & \small{-0.26179}  & \small{-0.26179}  & \small{-0.26179}   & \small{-0.26179} & \small{$-\pi /12$}  \\ \vspace{-0.15cm}
  \small{1}    &  \small{-0.75462}  & \small{-0.75488}  & \small{-0.75491} & \small{-0.75492}   & \small{-0.75493} & \small{-0.75494}    \\  \vspace{-0.15cm}
  \small{2}    &  \small{-1.25211}  & \small{-1.25287}  & \small{-1.25296} & \small{-1.25300}   & \small{-1.25301} & \small{-1.25306}     \\ \vspace{-0.15cm}
  \small{3}    &  \small{-1.75030}  & \small{-1.75183}  & \small{-1.75202} & \small{-1.75208}   & \small{-1.75211} & \small{-1.75221}   \\ \vspace{-0.15cm}
  \small{4}    &  \small{-2.24855}  & \small{-2.25109}  & \small{-2.25141} & \small{-2.25151}   & \small{-2.25156} &\small{-2.25172}\\ \vspace{-0.15cm}
  \small{5}    &  \small{-2.74666}  & \small{-2.75046}  & \small{-2.75094} & \small{-2.75110}   & \small{-2.75118} &\small{-2.75141}\\ \vspace{-0.15cm}
  \small{6}    &  \small{-3.24454}  & \small{-3.24986}  & \small{-3.25053} & \small{-3.25075}   & \small{-3.25086} &\small{-3.25120}\\ \vspace{-0.15cm}
  \small{7}    &  \small{-3.74217}  & \small{-3.74926}  & \small{-3.75015} & \small{-3.75044}   & \small{-3.75059} &\small{-3.75104}\\ \vspace{-0.15cm}
  \small{8}    &  \small{-4.23952}  & \small{-4.24863}  & \small{-4.24977} & \small{-4.25015}   & \small{-4.25034} &\small{-4.25092}\\ \vspace{-0.15cm}
  \small{9}    &  \small{-4.73658}  & \small{-4.74796}  & \small{-4.74939} & \small{-4.74987}   & \small{-4.75010} &\small{-4.75082}\\ \vspace{-0.15cm}
  \small{10}   &  \small{-5.23335}  & \small{-5.24725}  & \small{-5.24899} & \small{-5.24958}   & \small{-5.24987} &\small{-5.25074}\\ \vspace{-0.15cm}
  \small{11}   &  \small{-5.72982}  & \small{-5.74648}  & \small{-5.74858} & \small{-5.74928}   & \small{-5.74963} &\small{-5.75068}\\ \vspace{-0.15cm}
  \small{12}   &  \small{-6.22599}  & \small{-6.24567}  & \small{-6.24814} & \small{-6.24897}   & \small{-6.24938} &\small{-6.25062}\\ \vspace{-0.15cm}
  \small{13}   &  \small{-6.72186}  & \small{-6.74480}  & \small{-6.74768} & \small{-6.74865}   & \small{-6.74913} &\small{-6.75058}\\ \vspace{-0.15cm}
  \small{14}   &  \small{-7.21742}  & \small{-7.24387}  & \small{-7.24720} & \small{-7.24831}   & \small{-7.24887} &\small{-7.25054}\\ \vspace{-0.15cm}
  \small{15}   &  \small{-7.71268}  & \small{-7.74288}  & \small{-7.74669} & \small{-7.74796}   & \small{-7.74859} &\small{-7.75050}\\ \vspace{-0.15cm}
  \small{16}   &  \small{-8.20764}  & \small{-8.24183}  & \small{-8.24615} & \small{-8.24759}   & \small{-8.24831} &\small{-8.25047}\\ \vspace{-0.15cm}
  \small{17}   &  \small{-8.70229}  & \small{-8.74073}  & \small{-8.74558} & \small{-8.74720}   & \small{-8.74801} &\small{-8.75045}\\ \vspace{-0.15cm}
  \small{18}   &  \small{-9.19663}  & \small{-9.23956}  & \small{-9.24498} & \small{-9.24680}   & \small{-9.24770} &\small{-9.25042}\\ \vspace{-0.15cm}
  \small{19}   &  \small{-9.69067}  & \small{-9.73834}  & \small{-9.74436} & \small{-9.74637}   & \small{-9.74738} &\small{-9.75040}\\ \vspace{-0.15cm}
  \small{20}   &  \small{-10.18441}  & \small{-10.23705}  & \small{-10.24370}& \small{-10.24593}  & \small{-10.24704} &\small{-10.25038}\\
  \hspace{1mm}\small{21}   &  \small{-10.67784}  & \small{-10.73570}  & \small{-10.74302}& \small{-10.74546}  & \small{-10.74669} &\small{-10.75036}\\ \hline
\end{tabular}\caption{\label{tab.3} The results of numerical integrations are listed
   for various values of $\ell$ and the cutoff $M$. All of these values are denoted by $I(\ell,M)$.
   For the last column, we define $f(\nu)=|I(\ell,\infty)|$, where $\nu=\ell+\frac{1}{2}$}
  \label{geometry}
\end{table}
\par
It can be shown that the integral\,(\ref{final. cas.}) is an odd
function of $\nu$. An appropriate ansatz for the integral, which can
be obtained by fitting its asymptotic values reported in
Table\,(\ref{tab.3}), is
\begin{eqnarray}\label{func.suppo.}\begin{array}{ll}
  f(\nu)=A \nu+B \nu^{-1}+C
  \nu^{-3}+D\nu^{-5}+\mathcal{O}(\nu)^{-7},\hspace{0.5cm}  \mbox{where}\hspace{0.5cm} \nu=\ell+\frac{1}{2}
\end{array}\end{eqnarray}
where $f(\nu)$ indicates the absolute value of the asymptotic values
of our numerical integration\,($I(\ell,M)$) recorded in the last
column of Table\,(\ref{tab.3}), and $A$,$B$,$C$ and $D$ are the
unknown coefficients to be determined from the data.  To find these
coefficients we fit an increasing sequence of data points to the
above ansatz, and then find the overall asymptotic values of these
coefficients. To be more concrete, we fit the values of the
following set of sets of data points:
$\ell$=\,$\{\{1,2,3,4,5\},\{1,2,3,4,5,6\},...,\{1,2,3,...,21\}\}$,
and for each set\,(which contains $N$ data points)\,we find the
corresponding coefficients $A_{N}$, $B_{N}$,$C_{N}$ and $D_{N}$.
\par
\begin{figure}[th]
 \hspace{0.7cm} \includegraphics[width=4.9cm]{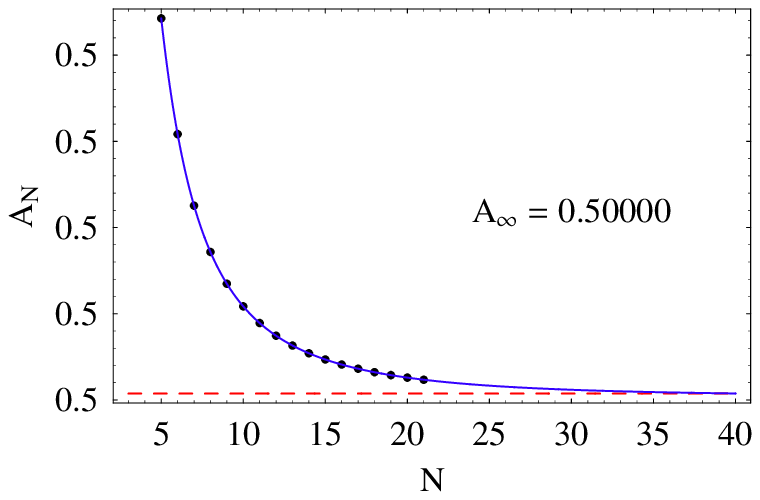}
 \hspace{0.3cm}  \includegraphics[width=5.4cm]{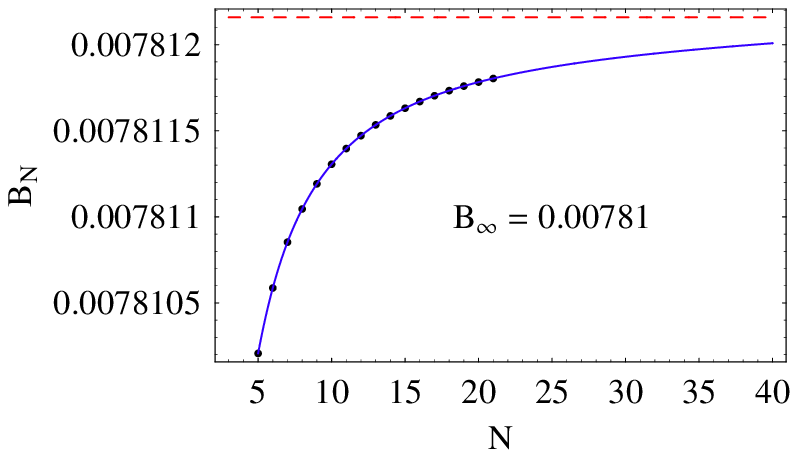}\\
     \hspace{0cm}\includegraphics[width=5.6cm]{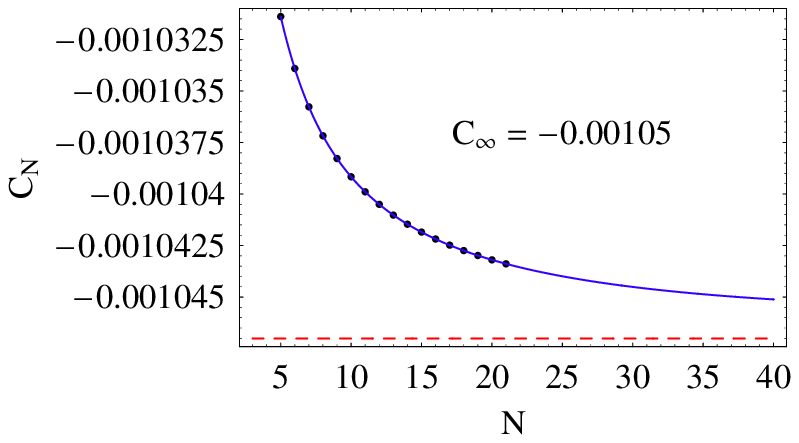}
      \hspace{0.5cm}   \includegraphics[width=5.4cm]{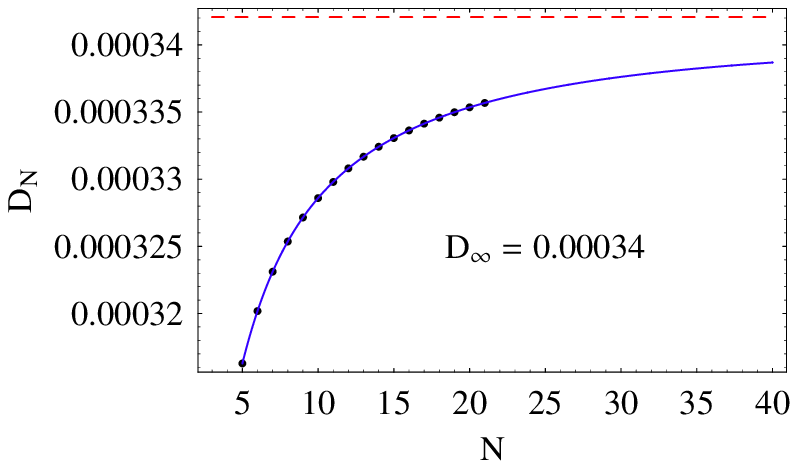}
\caption{\label{figs.asymp.c.n.} The plots of $A_{N}$,
$B_{N}$,$C_{N}$ and $D_{N}$ as a function of $N$, as explained in
the text. These four graphs clearly show that there are asymptotic
values for the coefficients which we denote by $A_{\infty}$,
$B_{\infty}$, $C_{\infty}$ and $D_{\infty}$.}
\end{figure}

In Fig.(\ref{figs.asymp.c.n.}) we plot $A_{N}$, $B_{N}$,$C_{N}$ and
$D_{N}$ as a function of $N$. These four graphs clearly show that
there are asymptotic values for the coefficients which we denote by
$A_{\infty}$, $B_{\infty}$, $C_{\infty}$ and $D_{\infty}$. We
had $17$ sets of data points for each graph, we fit a polynomial with terms
$N^{-j}$ with $j\in \{-5,-4,...,0\}$, and obtained the asymptotic
values,
\begin{eqnarray}\label{coefs.}\begin{array}{ll}\hspace{-2.2cm}
  A_{\infty}=0.50000\hspace{.5cm},\hspace{0.5cm}
  B_{\infty}=0.00781\hspace{.5cm},\hspace{0.5cm}
  C_{\infty}=-0.00105\hspace{.5cm},\hspace{0.5cm}D_{\infty}=0.00034,
\end{array}\end{eqnarray}
so we have,
\begin{eqnarray}\label{func.}\begin{array}{ll}
  f(\nu)=A_{\infty}\nu+B_{\infty}\nu^{-1}+C_{\infty}\nu^{-3}+D_{\infty}\nu^{-5}+\mathcal{O}(\nu)^{-7}\hspace{1cm}\small{\ell>0}.
\end{array}\end{eqnarray}
Before proceeding with our calculation, it is interesting to mention
that $f(\nu)$ has the following reported asymptotic expansion (Eq.\,(3.2) in Ref.\,\cite{nestrenko.}),
\begin{eqnarray}\label{func.analytic.}
  \small{f(\nu)=\frac{\nu}{2}+\frac{1}{128\nu}-\frac{35}{32768\nu^{3}}+\frac{565}{1048576\nu^{5}}+\mathcal{O}(\nu)^{-7}}\hspace{1cm}\small{\ell>0}.
\end{eqnarray}
Comparing these two expressions, it is obvious that they are
equivalent, to within the numerical accuracy of our calculation. Now
we can substitute our optimal functional form obtained for the
integral, Eq.\,(\ref{func.}), into Eq.\,(\ref{final. cas.}) to
obtain,
\begin{eqnarray}\label{zeta. func.}\hspace{-2cm}
  \footnotesize{E_{\mbox{\footnotesize Cas.}}=\frac{\hbar
  c}{2a}\sum\limits_{s=1}^{\infty}s\pi+\frac{-\hbar
  c}{2a}\bigg(\sum\limits_{\ell=1}^{\infty}2\nu f(\nu)\bigg)-\{a\to
  R/\eta\}}\nonumber\\ \hspace{-0.9cm}
  =\frac{\hbar c}{2a}\Bigg[\zeta(-1)\pi-
  2A_{\infty}\zeta(-2,3/2)-2B_{\infty}\zeta(0,3/2)-2C_{\infty}\zeta(2,3/2)\nonumber\\ \hspace{-0.6cm}-2D_{\infty}\zeta(4,3/2))\Bigg]-\{a\to
  R/\eta\}\nonumber\\ \hspace{-0.9cm}
  =\frac{\hbar c}{2a}\Bigg[\frac{-\pi}{12}-
  2A_{\infty}\bigg(\frac{-1}{4}\bigg)-2B_{\infty}\big(-1\big)-2C_{\infty}(\frac{\pi^{2}}{2}-4)
  \nonumber\\ \hspace{-0.6cm}-2D_{\infty}(\frac{\pi^{4}}{6}-16)\Bigg]-\{a\to
  R/\eta\},
\end{eqnarray}
where in the last step we have used the analytic continuations of
the generalized zeta
functions\,\cite{lecture.note.elizalde,Elizalde.book.1} along with
the following shift formula for the zeta functions,
\begin{eqnarray}\label{math.zeta.}\begin{array}{ll}\hspace{-2cm}
\zeta(n,m)=\sum\limits_{k=0}^{\infty}\left(k+m\right)^{-n}=m^{-n}+\sum\limits_{k'=0}^{\infty}\left(k'+m+1\right)^{-n}=m^{-n}+\zeta({n,m+1}).
\end{array}\end{eqnarray}
Obviously, in this case the last term in Eq.\,(\ref{zeta. func.})
goes to zero as $R\to \infty$. We have checked that we can obtain
the same finite result, without resorting to any analytic
continuation, using the Abel-Plana summation formula. We have
separated the $\ell=0$ term from the rest of the summation over
$\ell$, primarily because the value of the integral for $\ell=0$
dose not follow as closely as we like the pattern of other data
points. We have checked that our shift formula allows us to single
out this term, even when we need to analytically continue the zeta
functions.
\par
Our final result for the Casimir energy for a massless scalar field
with Dirichlet boundary condition in a three dimensional sphere is,
\begin{eqnarray}\label{final. result.}\hspace{2.5cm}
  E_{\mbox{\footnotesize Cas.}}=+0.00562\left(\frac{\hbar c}{2a}\right).
\end{eqnarray}
This result has been obtained numerically and is the same as the
previously reported result \,\cite{positive.,nestrenko.,Bowers.},
which were obtained by different numerical methods, to within
$0.2\%$.

\section{Conclusion}

In this paper, the Casimir energy for a massless scalar field in a
sphere with radius $a$ with Dirichlet boundary condition is computed
in three spatial dimensions. This energy has been computed using the
BSS which is based on confining the original sphere inside a
concentric larger sphere and computing the difference between the
vacuum energies of two configurations which differ only by the size
of the inner spheres. Finally, the radii of all the spheres except
the original one go to infinity. Our result shows that the Casimir
energy for a massless scalar field in a cube and a sphere have
indeed opposite signs, contrary to the case of the EM field. As
mentioned in the Introduction, and explicitly shown in
Table\,(\ref{tab.compare.}), the sign of the Casimir energy for
closed surfaces depends crucially on the type of field considered,
the boundary conditions imposed, and the details of the geometry and
not merely the topology. To be more specific, the claim that the
sign of the Casimir energy should not have a significant dependency
on the details of the shapes of the
vessels\,\cite{Lukosz.,wolf.,peterson.hansson.} is, much to our
surprise, not true.

\appendix
\section{The Imposition of the Boundary Conditions} \label{Calculation.1}
\setcounter{equation}{0}
\renewcommand{\theequation}{\Alph{section}.\arabic{equation}}

In this Appendix, we show how the imposition of Dirichlet boundary
condition on the inner and outer spheres relates the root indices,
$s$ and $S$, and the wave-vectors $k_{\ell,s}\left(a\right)$ and
$K_{\ell,S}\left(a,R\right)$, which appear in the main regulated
expression for the Casimir energy, Eq.\,(\ref{ECas.replace.}). For
the configuration A, we obtain the following condition for the inner
sphere
\begin{eqnarray}\label{boundary.con.in.}
  j_{\ell}\left(ak_{\ell,s}\left(a\right)\right)=0.
\end{eqnarray}
That is $k_{\ell,s}(a)$ is the $s$th zero of
$j_{\ell}(ak_{\ell,s}(a))$. The ranges of the values of angular
momentum and inner root indices are $\ell=\{0,1,2,3,...\}$ and
$s=\{1,2,3,...\}$. Similarly, the following relationship can be obtained
by combining the results of the imposition of the Dirichlet boundary
conditions on the two surfaces confining the annular region,
\begin{eqnarray}\label{boundary.con.out.}
  \dd\frac{j_{\ell}\left(aK_{\ell,S}\left(a,R\right)\right)}{n_{\ell}\left(aK_{\ell,S}\left(a,R\right)\right)}-
  \frac{j_{\ell}\left(RK_{\ell,S}\left(a,R\right)\right)}{n_{\ell}\left(RK_{\ell,S}\left(a,R\right)\right)}=0,
\end{eqnarray}
where $K_{\ell,S}(a,R)$ is the $S$th zero of
Eq.\,(\ref{boundary.con.out.}). The ranges of the values of angular
momentum and annular root indices are $\ell=\{0,1,2,3,...\}$ and
$S=\{1,2,3,...\}$. In order to find the relationship between the root
indices and the wave-vectors, we use the following relationships
between the spherical Bessel functions, trigonometric functions, and
polynomials in $x^{-1}$,
\begin{eqnarray}\label{expand.j&n}
  \begin{array}{l}\vspace{0.3cm}
  x j_{\ell}(x)=\sin\left(x-\frac{1}{2} \ell \pi\right)A_{\ell}(x)+\cos\left(x-\frac{1}{2} \ell \pi\right)B_{\ell}(x), \\
  x n_{\ell}(x)=-\cos\left(x-\frac{1}{2} \ell \pi\right)A_{\ell}(x)+\sin\left(x-\frac{1}{2} \ell \pi\right)B_{\ell}(x),
\end{array}\end{eqnarray}
where
\begin{eqnarray}\label{expand.coeff1.}\hspace{0cm}
  A_{\ell}(x)=\sum\limits_{r = 0}^{[\frac{1}{2}\ell\,]}\frac{\left(-1\right)^{r}(\ell+2r)!}{(2r)!(\ell-2r)!}\left(4 x^{2}\right)^{-r},
 \nonumber \\  B_{\ell}(x)=\frac{1}{2x}\sum\limits_{r = 0}^{[\frac{1}{2}\ell-\frac{1}{2}\,]}\frac{\left(-1\right)^{r}(\ell+2r+1)!}{(2r+1)!(\ell-2r-1)!}\left(4 x^{2}\right)^{-r}.
\end{eqnarray}
Using Eqs.\,(\ref{expand.j&n},\ref{expand.coeff1.}) and
Eq.\,(\ref{boundary.con.in.}), we obtain the following nonlinear
relation between integer $s$ and $k_{\ell,s}\left(a\right)$,
\begin{eqnarray}\label{inner. eq1.}
  ak_{\ell,s}\left(a\right)-\frac{1}{2}\ell \pi=\pi
  s-\arctan\left[\frac{B_{\ell}\left(ak_{\ell,s}\left(a\right)\right)}{A_{\ell}\left(ak_{\ell,s}\left(a\right)\right)}\right],
\end{eqnarray}
which can be simplify to,
\begin{eqnarray}\label{inner. eq2.}
  \pi s=-\arctan\left[\frac{j_{\ell}\left(ak_{\ell,s}\left(a\right)\right)}{n_{\ell}\left(ak_{\ell,s}\left(a\right)\right)}\right].
\end{eqnarray}
Similarly, we can repeat analogous steps to obtain an equation for
integer $S$ as a function of $K_{\ell,S}(a,R)$ for the annular
region,
\begin{eqnarray}\label{annular. eq.}
  \pi S=\arctan\left[\frac{j_{\ell}\left(aK_{\ell,S}\left(a,R\right)\right)}{n_{\ell}\left(aK_{\ell,S}\left(a,R\right)\right)}\right]
  -\arctan\left[\frac{j_{\ell}\left(RK_{\ell,S}\left(a,R\right)\right)}{n_{\ell}\left(RK_{\ell,S}\left(a,R\right)\right)}\right].
\end{eqnarray}
Comparing Eq.\,(\ref{inner. eq2.}) with Eq.\,(\ref{annular. eq.}) we
obtain the following relationship between the root indices,
\begin{eqnarray}\label{s and S eqs.}
  S_{\ell}\left(a,R,K\right)=s_{\ell}\left(R,K\right)-s_{\ell}\left(a,K\right).
\end{eqnarray}
Note that, in this equation the values of the root indices $s$, $S$,
and the wave-vector $K$ can be thought of as having been
analytically continued to non-integer values.

\section{Implementing the $R\to\infty$ Limit} \label{Calculation}

\setcounter{equation}{0}
\renewcommand{\theequation}{\Alph{section}.\arabic{equation}}

In this Appendix, we show that in the limit $R\to \infty$, the
contribution to the Casimir energy, reflected by the last two terms
in Eq.\,(\ref{ECas.replace.}), can be greatly simplified. This is
due to the fact that, as we shall show, most of the infinities
coming from the annular regions cancel due to BSS, and all of the
remaining finite terms approach zero. Using the EMSF, the
contribution of the annular region in Eq.\,(\ref{ECas.replace.}) is,
\begin{eqnarray}\label{elimination. R.1}\begin{array}{ll}\hspace{-3cm}\vspace{0mm}
  \dd\sum\limits_{S=1}^{\infty}\left[K_{\ell,S}\left(a,R\right)\,g\left(\lambda K_{\ell,S}\left(a,R\right)\right)
  -K_{\ell,S}\left(R/\eta,R\right)\,g\left(\lambda K_{\ell,S}\left(R/\eta,R\right)\right)\right]
  \\ \hspace{-2cm}\vspace{2mm}
  =\dd\int_{S=1}^{\infty}dS\left[K_{\ell}\left(a,R,S\right)\,g\left(\lambda K_{\ell}\left(a,R,S\right)\right)
  -K_{\ell}\left(R/\eta,R,S\right)\,g\left(\lambda K_{\ell}\left(R/\eta,R,S\right)\right)\right]
  \\ \hspace{-2cm}\vspace{2mm}+
  \dd\frac{1}{2}\left[K_{\ell,1}\left(a,R\right)\,g\left(\lambda K_{\ell,1}\left(a,R\right)\right)
  -K_{\ell,1}\left(R/\eta,R\right)\,g\left(\lambda K_{\ell,1}\left(R/\eta,R\right)\right)\right]
  \\ \hspace{-2cm}\vspace{2mm}
  \dd+\sum\limits_{r=1}^{r=N}\frac{B_{2r}}{(2r)!}\left\{\frac{d^{2r-1}}{dS^{2r-1}}\Bigg[K_{\ell}(a,R,S)g(\lambda
  K_{\ell}(a,R,S))-K_{\ell}(R/\eta,R,S)g(\lambda
  K_{\ell}(R/\eta,R,S))\Bigg]\right\}_{S=1}\\ \hspace{-2cm}\vspace{2mm}
  \dd-\int_{S=1}^{\infty}dS\,B_{2N}\left(S-[S]\right)\\ \hspace{-.8cm}\dd\times\frac{d^{2N}}{dS^{2N}}
  \Bigg[K_{\ell}\left(a,R,S\right)\,g\left(\lambda K_{\ell}\left(a,R,S\right)\right)
  -K_{\ell}\left(R/\eta,R,S\right)\,g\left(\lambda K_{\ell}\left(R/\eta,R,S\right)\right)\Bigg].
\end{array}\end{eqnarray}
Here we show that the last three terms in the right hand side\,(rhs)
of Eq.\,(\ref{elimination. R.1}) give no contribution to the Casimir
energy. In fact, in the limit $R\to\infty$, all terms in the rhs of
Eq.\,(\ref{elimination. R.1}), except for the first integral, cancel
each other due to BSS or vanish entirely. In order to show the
vanishing of these terms, we have to prove two things. First, each
of these terms decreases rapidly as a function of $\ell$, so that
when it is summed over $\ell$ with a pre-factor of $2\ell+1$ it will
not give a divergent contribution. Second, each term approaches zero
in the limit $R\to \infty$.
\par
We start with the second term. For the small
values of $\ell$, the first root goes to zero as $R^{-1}$. For large
values of $\ell$, the first root $K_{\ell,1}(R/\eta,R)$ behaves
like,
\begin{eqnarray}\label{K(a)-K(R/eta.1}
  K_{\ell,1}(R/\eta,R)=k_{\ell,1}(R)+\frac{\epsilon}{R}
\end{eqnarray}
where
\begin{eqnarray}\label{K(a)-K(R/eta.2}\hspace{-1cm}
  \epsilon\sim\frac{\nu ^{1/3}}{1.23\eta}e^{-2\nu (\beta_{\nu}-\tanh
  \beta_{\nu})}\hspace{1cm}\textmd{,
  and}\hspace{1cm}(\cosh\beta_{\nu})^{-1}\sim\eta^{-1}[1+\mathcal{O}(\nu^{-2/3})].
\end{eqnarray}
and $\nu=\ell+1/2$. Therefore, the second term becomes,
\begin{eqnarray}\label{K(a)-K(R/eta.3}
  \hspace{-2.5cm}\frac{1}{2}\Bigg[
  \left(k_{\ell,1}(R)+\frac{\epsilon}{R}\right)e^{-\lambda
  \left(k_{\ell,1}(R)+\frac{\epsilon}{R}\right)}-\left(k_{\ell,1}(R)+\frac{\epsilon'}{R}\right)e^{-\lambda
  \left(k_{\ell,1}(R)+\frac{\epsilon'}{R}\right)}\Bigg]\nonumber\\ \hspace{-0.7cm}
  =\frac{1}{2}\Bigg[\left(k_{\ell,1}(R)+\frac{\epsilon}{R}\right)-\lambda
  \left(k_{\ell,1}(R)+\frac{\epsilon}{R}\right)^{2}+...\\ \hspace{0.2cm}-\left(k_{\ell,1}(R)+\frac{\epsilon'}{R}\right)+\lambda
  \left(k_{\ell,1}(R)+\frac{\epsilon'}{R}\right)^{2}+...\Bigg]
  \sim\mathcal{O}(\epsilon/R)-\mathcal{O}(\epsilon'/R),\nonumber
\end{eqnarray}
For obtaining the second line we have expanded the exponentials
since $\lambda$ eventually approaches zero. As shown in
Eq.\,(\ref{K(a)-K(R/eta.2}) the numerators $\epsilon$ and
$\epsilon'$ decrease exponentially in $\ell$, and the entire term
goes to zero in the limit $R \to \infty$.
\par An analogous argument can be used for the third term in the rhs
of Eq.\,(\ref{elimination. R.1}). In the neighborhood of $S=1$, we
have the following relationship between the wave-vectors,
\begin{eqnarray}\label{expand.k.any S.}
  K_{\ell}(a,R,S)=k_{\ell}(R,S)+\delta_{\ell}(a,R,S).
\end{eqnarray}
where $\delta_{\ell}(a,R,S)$ decreases exponentially with increasing
$\ell$ and vanishes as $R\to \infty$. Analogous relationship holds
for any derivatives of the wave-vectors. In fact, all terms which
depend only upon the outer radius $R$\,(and not on the inner
ones\,$a$ or $R/\eta$) are canceled due to BSS, and the remaining
terms go to zero as $R\to \infty$.
\par Now, we proceed to the fourth term. Here, we show that the remaining integral decreases rapidly with increasing
$\ell$. We rewrite the integral as
\begin{eqnarray}\label{elimination. R.2}
  \hspace{-2.5cm}\Bigg{|}\int_{S=1}^{\infty}dS\,B_{2N}\left(S-[S]\right)\frac{d^{2N}}{dS^{2N}}
  \Bigg[K_{\ell}\left(a,R,S\right)\,g\left(\lambda
  K_{\ell}\left(a,R,S\right)\right)-\{a\to R/\eta\}
  \Bigg]\Bigg{|}\nonumber
  \\  \hspace{-1cm}\leq \frac{|B_{2N}|}{(2N)!}
  \int_{S=1}^{S=S^{*}} dS\left| \frac{d^{2N}}{dS^{2N}}
  \Bigg[K_{\ell}\left(a,R,S\right)\,g\left(\lambda
  K_{\ell}\left(a,R,S\right)\right)-\{a\to R/\eta\}\Bigg]\right|\nonumber\\
  \hspace{-0.85cm}+\frac{|B_{2N}|}{(2N)!} \int_{S=S^{*}}^{S=\infty} dS
  \left|\frac{d^{2N}}{dS^{2N}}\left[K_{\ell}\left(a,R,S\right)\,g\left(\lambda
  K_{\ell}\left(a,R,S\right)\right)\right]\right|\nonumber\\
  \hspace{-0.85cm}+\frac{|B_{2N}|}{(2N)!} \int_{S=S^{*}}^{S=\infty} dS
  \left|\frac{d^{2N}}{dS^{2N}}\left[K_{\ell}\left(R/\eta,R,S\right)\,g\left(\lambda
  K_{\ell}\left(R/\eta,R,S\right)\right)\right]\right|
\end{eqnarray}
where $S^{*}=S_{\ell}(R/\eta,R,\nu/b)$ and $b$ is constrained
by:\,$R/\eta<b<R$. For large values of $\ell$, we can repeat
analogous arguments given above to show that for the first integral
the terms of the integrand which depend only upon the outer radius
$R$, are canceled due to BSS, and the result of the integration of
the remaining terms decrease exponentially with increasing $\ell$
and vanish for $R\to \infty$. As for the second term in the rhs of
Eq.\,(\ref{elimination. R.2}), the high order derivatives of the
wave-vector $K_{\ell}\left(a,R,S\right)$ for large value of $\ell$
are,
\begin{eqnarray}\label{high.derivatives.K}
\frac{d^{m}K_{\ell}\left(a,R,S\right)}{dS^{m}}&=&\mathcal{O}({\nu}^{1-m}),\hspace{0.5cm}\mbox{for}\hspace{0.5cm}\nu/R\ll
K_{\ell}\left(a,R,S\right)\ll
\nu/a\nonumber\\&=&\mathcal{O}(\nu^{-m/3}),\hspace{0.35cm}\mbox{for}\hspace{0.5cm}K_{\ell}\left(a,R,S\right)\sim(\nu+\tau
\nu^{1/3})/a\nonumber\\&=&\mathcal{O}(\nu^{1-m}),\hspace{0.5cm}\mbox{for}\hspace{0.5cm}\nu/a\ll
K_{\ell}\left(a,R,S\right)
\end{eqnarray}
where $\nu=\ell+1/2$ and $m>1$. Thus if a sufficient number of terms
are taken in the EMSF (Eq.\,(\ref{elimination. R.1})) the high order
derivatives of the wave-vector, $K_{\ell}$, decrease with increasing
$\nu$. Moreover, when derivatives operate on the damping factors,
$g(\lambda K_{\ell}\left(a,R,S\right))=e^{-\lambda
K_{\ell}\left(a,R,S\right)}$, each derivative extracts a factor of
$\lambda$,
\begin{eqnarray}\label{high.derivatives.g}
\frac{d}{dS}g(\lambda
K_{\ell}\left(a,R,S\right))\sim\lambda\mathcal{O}\Big[\frac{1}{R}g(\lambda
K_{\ell}\left(a,R,S\right))\Big].
\end{eqnarray}
Now, if $N$ in Eq.\,(\ref{elimination. R.1}) is chosen to be large
enough, each term arising from
$\left|\frac{d^{2N}}{dS^{2N}}\left[K_{\ell}\left(a,R,S\right)\,g\left(\lambda
  K_{\ell}\left(a,R,S\right)\right)\right]\right|$, decreases rapidly with increasing $\nu$ or has many factors of
$\lambda$. The terms which have factors of $\lambda$ approach zero
when $\lambda$ goes to the zero, and for any remaining terms, their
integral decrease rapidly as a function of $\nu$ and the
contribution of these terms approach zero in the limit $R\to
\infty$. Analogous argument can be repeated for the last integral in
the rhs of Eq.\,(\ref{elimination. R.2}), only with the replacement
$a\to R/\eta$. Therefore, we have shown that the left hand side of
Eq.\,(\ref{elimination. R.2}) approaches zero. Finally, the result
of this analysis is that the only contribution coming from the
annular region is the integral term in the rhs of
Eq.\,(\ref{elimination. R.1}). Therefore, the summation in the
Eq.\,(\ref{ECas.replace.}) can be replaced by its integral.

\section*{Acknowledgement} We would like to thank the research office
of the Shahid Beheshti University for financial support.

 \end{document}